\documentclass[journal,twoside,web]{ieeecolor}
\usepackage{generic}
\usepackage{cite}
\usepackage{amsmath,amssymb,amsfonts}
\usepackage{graphicx}
\usepackage{algorithm,algorithmic}
\usepackage{hyperref}
\hypersetup{hidelinks}
\usepackage{textcomp}
\usepackage[utf8]{inputenc}
\usepackage{changepage}
\usepackage{graphics,graphicx}
\usepackage{mathrsfs} 
\usepackage{comment}
\usepackage{algorithm}
\usepackage{algorithmic}
\usepackage[linesnumbered,ruled,vlined,algo2e]{algorithm2e}
\usepackage{xcolor}
\usepackage{comment}
\usepackage{bm}
\usepackage{bbm}
\usepackage{enumerate}
\usepackage{commath}

\usepackage{subcaption}

\usepackage{cite}
\usepackage{booktabs}
\usepackage{empheq}
\usepackage{amssymb}

\usepackage{amsthm}
\usepackage{accents}
\usepackage{thmtools}
\usepackage{thm-restate}
\usepackage{float}
\usepackage[normalem]{ulem}

\theoremstyle{plain}

\newtheorem{lemma}{Lemma}

\newtheorem{theorem}{Theorem}

\newtheorem{problem}{Problem}
\newtheorem*{problem*}{Problem}
\newtheorem*{theorem*}{Theorem}
\newtheorem{assumption*}{Assumption}

\newtheorem{remark}{Remark}

\theoremstyle{definition}
\newtheorem{example}{Example}

\def\BibTeX{{\rm B\kern-.05em{\sc i\kern-.025em b}\kern-.08em
    T\kern-.1667em\lower.7ex\hbox{E}\kern-.125emX}}

\begin{document}

\title{Estimating unknown dynamics and cost as a bilinear system with Koopman-based Inverse Optimal Control}

\author{Victor Nan Fernandez-Ayala$^{1}$, \IEEEmembership{Graduate Student, IEEE}, Shankar A. Deka$^{2}$, \IEEEmembership{Member, IEEE}, and Dimos V. Dimarogonas$^{1}$, \IEEEmembership{Fellow, IEEE}
\thanks{*This work was supported by the the ERC CoG LEAFHOUND, the EU CANOPIES project, the Knut and Alice Wallenberg Foundation (KAW) and the Digital Futures SHARCEX and Smart Construction projects.}
\thanks{$^{1}$Victor Nan Fernandez-Ayala and Dimos V. Dimarogonas are with the Division of Decision and Control Systems, School of EECS, Royal Institute of Technology (KTH), 100 44 Stockholm, Sweden (Email: 
        {\tt\small vnfa, dimos@kth.se}).}%
\thanks{$^{2}$Shankar A. Deka is with the Department of Electrical Engineering and Automation, School of Electrical Engineering, Aalto University, 02150 Espoo, Finland (Email: 
        {\tt\small shankar.deka@aalto.fi}).}%
}

\maketitle


\begin{abstract}
In this work, we address the challenge of approximating unknown system dynamics and cost functions through a Koopman-based Inverse Optimal Control (IOC) framework. Using optimal trajectories, a modified Extended Dynamic Mode Decomposition with control (EDMDc) constructs a bilinear control system in lifted coordinates. Pontryagin’s Maximum Principle (PMP) conditions are then derived, revealing structural similarities to the inverse Linear Quadratic Regulator (LQR) problem. This allows tractable cost recovery without resorting to nonlinear IOC formulations. The bilinear representation also inherits the analytical advantages of linear systems. Simulation and robotic experiments validate the approach, showing accurate estimation of both dynamics and costs, and illustrating its potential for general control and modeling applications.
\end{abstract}


\section{Introduction}
\label{sec:introduction}
\IEEEPARstart{E}{stimation} of unknown system dynamics and cost functions is a fundamental challenge in control and optimization theory. Powerful methods to solve subsets of this problem already exist, e.g., Sparse Identification of Nonlinear Dynamics (SINDy) \cite{sindy} which identifies nonlinear dynamics using a library of basis functions. One particularly difficult problem here is human motion prediction, which is crucial for applications such as human-robot interaction, autonomous vehicles, and rehabilitation robotics. Although some probabilistic models exist \cite{hil_probabilistic_model}, they are often simple and may not capture the full complexity of human motion, whose intricate dynamics and control strategies are inherently nonlinear.

In addition to estimating system dynamics, computing unknown cost functions is equally challenging. Inverse Optimal Control (IOC) \cite{distributed_ioc} and Inverse Reinforcement Learning (IRL) \cite{hil_irl} are methodologies developed to recover cost functions from observed behavior (e.g., state trajectories and control inputs). In the context of human motion, IOC can be particularly useful for predicting movements by inferring the cost functions that humans implicitly optimize. This approach is supported by neuroscience studies suggesting that human motor control follows the principles of optimality, such as minimizing effort or jerk \cite{Todorov2002}. Several methods have been proposed to estimate unknown dynamics and cost functions. The inverse Linear Quadratic Regulator (LQR) problem \cite{discrete_ioc} provides a framework for estimating cost functions when the system dynamics are linear and known. However, this approach cannot handle the nonlinear dynamics present in many real-world applications. To extend inverse LQR to nonlinear systems, some works have considered using Koopman operator theory to linearize the dynamics, e.g., \cite{koopman_ioc} which relies on deep neural networks to approximate the operator. Nevertheless, these methods can be complex and lack rigorous convergence analysis. Nonlinear IOC methods \cite{nonlinear_ioc} attempt to directly estimate cost functions for nonlinear systems, but often face challenges due to its complexity and the requirement of known dynamics, which limits their applicability when models are not readily available. Therefore, there is a need for methods that can simultaneously estimate unknown dynamics and cost functions.

In this paper, we propose a novel approach to learn unknown system dynamics and cost functions by leveraging bilinear Koopman representations for IOC. Specifically, a modified Extended Dynamic Mode Decomposition with control (EDMDc) \cite{koopman_brunton} is used to obtain a bilinear control system in lifted coordinates. Building on this model, Pontryagin’s Maximum Principle (PMP) conditions are derived, which resemble the inverse LQR problem due to state-control separability, and enable a tractable solution for estimating unknown cost functions without resorting to nonlinear deep learning-based Koopman methods. Compared to inverse LQR approaches \cite{discrete_ioc}, our method also handles nonlinear dynamics through the bilinear representation, while benefiting from the well-studied theory of bilinear control systems \cite{BilinearSystems}. We demonstrate the effectiveness of our method through simulations and a robotic experiment, highlighting its potential for broader applications in control, particularly in human motion prediction. To our knowledge, this is the first work to directly integrate Koopman lifting with IOC, in contrast to \cite{koopman_ioc} where Koopman is used separately for system identification followed by nonlinear IOC.

\section{Preliminaries}
Consider a control-affine system in discrete time given by
\begin{equation}\label{eq:disc_dynamics}
    x_{k+1}  = f(x_k) + g(x_k)u_k,
\end{equation}
\noindent where $x_k\in X\subset  \mathbb{R}^n,\;u_k\in U \subset \mathbb{R}^m$; and $f: X \rightarrow \mathbb{R}^n$ and $g: X \rightarrow \mathbb{R}^{n \times m}$ are twice continuously differentiable functions, i.e., $f, g \in \mathcal{C}^2(X)$, and the state space $X \subset \mathbb{R}^n$ and control input space $U \subset \mathbb{R}^m$ are compact sets. 

Let us denote the trajectory and control input sequence generated by \eqref{eq:disc_dynamics} in compact form as $x_{0:T} = \left\{x_0,x_1,\dots,x_T\right\}$ and $u_{0:T} = \left\{u_0,u_1,\dots,u_T\right\}$. We assume that \eqref{eq:disc_dynamics} is locally controllable on $X$, i.e., for any initial state $x_0 \in X$ and target state $x_T \in X$ there exists a control input sequence ${u}_{0:T-1}$ that steers the state from $x_0$ to $x_T$ within a finite time $T$.

Next, we consider cost functions of the form
\begin{equation}\label{eq:cost}
    J\left(x_{0:T},u_{0:T}\right) = \Phi(x_T) + \frac{1}{2} \sum_{k=0}^{T-1} l(x_k,u_k),
\end{equation}
where $l \in \mathcal{C}^2(X,U): X \times U \rightarrow \mathbb{R}$ is the stage cost, and $\Phi \in \mathcal{C}^2(X): X \rightarrow \mathbb{R}$ is the terminal cost. The optimal control objective is to minimize this cost function $J$, and the corresponding optimal trajectory and control sequence is denoted by 
$(x^*_{0:T}, u^*_{0:T}) = \arg \min J\left(x_{0:T},u_{0:T}\right)$. Given the system dynamics \eqref{eq:disc_dynamics} and $M$ optimal trajectory and control sequences $x^*_{0:T}, u^*_{0:T}$, the goal of IOC is to estimate the corresponding cost function \eqref{eq:cost}. We start by restricting the candidate cost functions to a finite dimensional space. We do this by parameterizing the stage cost $l(x,u)$ as follows
\begin{equation}\label{eq:param_cost}
    l_\omega(x,u) \doteq \sum_{i=1}^N \omega_{i}\phi_i(x,u), \quad J_\omega = \Phi(x_T) + \frac{1}{2} \sum_{k=0}^{T-1} l_\omega(x_k,u_k),
\end{equation}
where parameters $\omega_i \in \mathbb{R}$ and basis functions $\phi_i \in \mathcal{C}^2(X,U)$. Thus, IOC seeks to estimate a cost parameter vector $\hat{w} = [\hat{w}_1, \hat{w}_2, \ldots, \hat{w}_N]^\top $ such that

\[
    J_{\hat{\omega}}(x^*_{0:T},u^*_{0:T}) \le J_{\hat{\omega}}(x_{0:T},u_{0:T}),
\]
for all trajectories with a fixed starting point $x_0 = x^*_0$. We assume the parameterized cost $l_\omega$ is convex in $u$, ensuring a unique optimal control for each $x$, i.e., $\frac{\partial^2}{\partial u^2} \big[\sum_i \omega_i \phi_i(x,u)\big] \succ 0$, where ($\succ 0$) denotes that the left term is positive-definite.

In this work, we explore a bilinearization of the IOC problem that also allow us to handle unknown system dynamics. To achieve this, we utilize Koopman operator theory and EDMDc to approximate the unknown dynamics and reformulate the IOC problem in an easy to analyze, tractable manner.

\subsection{Koopman operator theory}
The Koopman operator \cite{koopman_brunton} provides a linear framework for analyzing nonlinear dynamical systems by lifting the state space to a linear higher-dimensional function space. For a discrete-time autonomous system $x_{k+1} = F(x_k)$, the Koopman semigroup of operators $\mathcal{K}$ acting on a Banach space of observables  $\psi: X \rightarrow \mathbb{C}$ is defined as $(\mathcal{K} \psi)(x) = \psi(F(x))$. This operator advances the observable functions along the trajectories of the system. Since it is infinite-dimensional, in practice, we approximate $\mathcal{K}$ using finite-dimensional methods, like Extended Dynamic Mode Decomposition (EDMD).

\subsubsection{Extended Dynamic Mode Decomposition with control} \label{sec:edmdc}
EDMDc extends the EDMD algorithm to control systems, enabling the approximation of the Koopman operator in systems with control inputs \cite{koopman_brunton}. We start by considering the usual case where the dynamics \eqref{eq:disc_dynamics} are linearized by converting the original state-space $x_k \in X$ into a lifted Koopman space $z_k = \theta(x_k) \in Z \subset \mathbb{R}^N$, wherein the states evolve as
\begin{equation} \label{eq:koopman_linear_disc}
    z_{k+1} = A z_k + B u_k, \quad x_k = C z_k, 
\end{equation}
and $A \in \mathbb{R}^{N \times N}$, $B \in \mathbb{R}^{N \times m}$ and $C \in \mathbb{R}^{n \times N}$ are the matrices to be determined using EDMDc. The function $\theta(x_k) = [\theta_1(x_k), \theta_2(x_k), \dots, \theta_N(x_k)]^\top$ is comprised of basis functions $\theta_i(x_k) \in \mathcal{F}, i = 1, 2, \dots,N$, where $\mathcal{F}$ is a Banach space.

Using $T$ optimal trajectory snapshots taken at some uniform sampling time $\Delta t$, in the form of pairs $(x_k, u_k, x_{k+1})$ where $x_{k+1}$ follows \eqref{eq:disc_dynamics} for $k = 0, 1, \dots,T-1$, we can estimate a finite dimensional approximation of the Koopman linear operator $\mathcal{K}$ restricted to the space spanned by the basis functions. The Koopman matrices $A$ and $B$ are then obtained by minimizing the residual error over the collected snapshots:
\begin{equation}\label{eq:koopman_approx}
\begin{aligned}
   & \min_{A,B} \sum_{k=0}^{T-1} \left| z_{k+1} - A z_k - B u_k \right|^2 
    = \\
    & \min_{\left[\begin{array}{cc} A & B \end{array}\right]} 
    \left\| \:
    \theta(Y)-\left[
    \begin{array}{cc} A & B \end{array} 
    \right] \left[
    \begin{array}{c} \theta(X) \\ \Upsilon
    \end{array}
    \right] \right\|_F^2,
\end{aligned}
\end{equation}
where $\| \cdot \|_F$ denotes the Frobenius norm, and matrices $\theta(X) = [\theta(x_0), \theta(x_1), \dots, \theta(x_{T-1})]$, $\Upsilon = [u_0, u_1, \dots, u_{T-1}]$ and $\theta(Y) = [\theta(x_1), \theta(x_2), \dots, \theta(x_{T})]$. The closed form solution is then given as
\begin{equation}\label{eq:koopman_approx_cf}
    \left[\begin{array}{cc} A & B \end{array}\right] = \theta(Y) \left[ \begin{array}{c} \theta(X) \\ \Upsilon \end{array} \right]^\dag,
\end{equation}
where $\dag$ denotes the pseudo-inverse. 

\begin{figure*}[hbt!]
\noindent\begin{minipage}{0.37\textwidth}
\begin{equation*}
\boxed{
\begin{aligned}
    &\textbf{Dynamics: } z_{k+1} = A z_k+B\left(u_k \otimes z_k\right) \\
    & \textbf{Unknown cost: }\frac{1}{2}\sum_{k=0}^{T-1} \bigg[ z_k^\top Q z_k + u_k^\top R u_k \bigg] \\
    & \textbf{Data available: } x^*_{0:T},u^*_{0:T} (\implies z_{0:T}^*,u^*_{0:T})
\end{aligned}}
\end{equation*}
\centering
(a) \textbf{Solve using: } Inverse (Bi-)LQR.
\end{minipage}
\begin{minipage}{0.12\textwidth}\centering
$\begin{array}{c}
    \text{\small Koopman}\\ \text{\small bi-linearization}\\
    \vspace{-0.3cm}\\
\mathbf{\longleftrightarrow}\normalfont
\end{array}$
\end{minipage}
\begin{minipage}{0.45\textwidth}
\begin{equation*}
\boxed{
\begin{aligned}
    &\textbf{Dynamics: } x_{k+1} = CA \theta(x_k) + C\left(u_k \otimes \theta(x_k)\right)\\
    &\textbf{Unknown cost: } \frac{1}{2} \sum_{k=0}^{T-1} \bigg[\sum_{i,j} q_{ij}\theta_i(x_k)^\top\theta_j(x_k) + r_{ij}u_{ik}u_{jk}\bigg]\\
    & \textbf{Data available: } x_{0:T}^*,u^*_{0:T} 
\end{aligned}}
\end{equation*}
\centering
(b) \textbf{Solve using: } Nonlinear IOC \cite{nonlinear_ioc}.
\end{minipage}
\caption{\textbf{Comparison between inverse (Bi-)LQR and nonlinear IOC approaches.} Unlike prior approaches, this framework uses a common parameterization for both dynamics and cost, enabling IOC even with unknown nonlinear dynamics.}
\label{fig:ioc_comparison}
\end{figure*}

\subsection{Inverse Optimal Control} \label{sec:ioc}
To apply IOC, we use the discrete-time PMP to write the necessary conditions satisfied by the optimal trajectory and input sequence $(x^*_{0:T},u^*_{0:T})$ using the Hamiltonian dynamics:
\begin{equation}\label{eq:PMP_condition_1}
    x^*_{k+1} = \frac{\partial}{\partial \lambda} H(x^*_k,\lambda_k,u^*_k),
\end{equation}
\begin{equation}\label{eq:PMP_condition_2}
    \lambda_{k} = \frac{\partial}{\partial x} H(x^*_k,\lambda_{k+1},u^*_k),
\end{equation}
\begin{equation}\label{eq:pmp_optimal} 
u_k^* = \arg \min_{u \in U} H(x_k, \lambda_{k+1}, u), 
\end{equation}
with terminal condition $\lambda_T = \frac{\partial}{\partial x}\Phi(x^*_T)$, where $\lambda_k$ is the co-state, and the Hamiltonian function $H(x,\lambda,u)$ is defined as $H \doteq \lambda^\top (f(x)+g(x)u) + l(x,u)$. Following \cite{nonlinear_ioc}, one can eliminate the co-state variables to obtain linear equations in the unknown parameters $\omega$ from \eqref{eq:param_cost}. Alternatively, we can reformulate the IOC problem using Koopman and parameterize the cost function $J$ such that we obtain an LQR cost as
\begin{equation} \label{eq:lqr_cost}
\begin{aligned}
    J(x_{0:T}, u_{0:T}) =  \frac{1}{2} \sum_{k=0}^{T-1} \bigg[ z_k^\top Q z_k + u_k^\top R u_k \bigg] = \\ \frac{1}{2} \sum_{k=0}^{T-1} \bigg[ \theta(x_k)^\top Q \theta(x_k) + u_k^\top R u_k \bigg],
\end{aligned}
\end{equation}
for some matrices $Q \succ 0_{N \times N}$ and $R \succ 0_{m \times m}.$ We note that this LQR cost function in lifted states $z$ can be ultimately parameterized in the form of equation \eqref{eq:param_cost} (as shown in Fig.~\ref{fig:ioc_comparison}) and that, for simplicity, we assume the final cost $\Phi(x_T)=0$.

\subsubsection{Inverse Linear-Quadratic Regulator} \label{sec:ilqr}
The ``forward" LQR problem with linear dynamics in the lifted states $z$ becomes
\begin{equation}\label{eq:lqr_lifted}
    \min_{x_{0:T},u_{0:T-1}} J = \frac{1}{2} \sum_{k=0}^{T-1} \bigg[ z_k^\top Q z_k + u_k^\top R u_k \bigg],
\end{equation}
\begin{equation}\label{eq:lifted_dynamics}
    s.t. \ z_{k+1} = A z_k + B u_k.
\end{equation}

Applying the PMP conditions \eqref{eq:PMP_condition_1}, \eqref{eq:PMP_condition_2} to the optimization problem \eqref{eq:lqr_lifted} and dynamics \eqref{eq:lifted_dynamics} gives the following conditions
\begin{eqnarray}
    \lambda_{k} &=& A^\top \lambda_{k+1} + Q z_k, \; k = 1:T-1,\\
    \lambda_{T} &=& 0, \\
    u_k &=& -B^\top \lambda_{k+1}, \; k = 0:T-1, \label{eq:ilqr_control}
\end{eqnarray}
where $R=I$ for simplicity. As described in \cite{discrete_ioc}, these equations can be re-arranged to create a linear system with separated variables of interest as the unknown (the $Q$ matrix)
\begin{equation}\label{eq:equation_of_interest}
    -\operatorname{vec}\left(u_{0: T-2}^{(1: M)}\right) = \mathscr{A}(z) \operatorname{vec}(Q) = \mathscr{A}(z) \mathscr{D} \operatorname{vech}(Q),
\end{equation}
where $\operatorname{vec}(\cdot)$ and $\operatorname{vech}(\cdot)$ denote the vectorization and half-vectorization, $u_{0: T-2}^{(1: M)}$ and $z_{0: T-1}^{(1: M)}$ are the stacked vector of all optimal trajectories and controls for $M$ different optimal trajectories, $\mathscr{D}$ is the duplication matrix and $\mathscr{A}(z)$ is 
\begin{equation}\label{eq:AA}
\begin{aligned}
\mathscr{A}(z)=\left[\begin{array}{c}
    z_1^{(1)^\top} \\
    \vdots \\
    z_1^{(M)^\top}
\end{array}\right] 
\otimes \left[\begin{array}{c}
    B^\top \\
    0 \\
    \vdots \\
    0
\end{array}\right] + \left[\begin{array}{c}
    z_2^{(1)^\top} \\
    \vdots \\
    z_2^{(M)^\top}
\end{array}\right] \otimes \\ \left[\begin{array}{c}
    B^\top A^\top \\
    B^\top \\
    \vdots \\
    0
\end{array}\right] + \cdots + \left[\begin{array}{c}
    z_{T-1}^{(1)^\top} \\
    \vdots \\
    z_{T-1}^{(M)^\top}
\end{array}\right] \otimes\left[\begin{array}{c}
    B^\top\left(A^\top\right)^{T-3} \\
    B^\top\left(A^\top\right)^{T-4} \\
    \vdots \\
    B^\top
\end{array}\right],
\end{aligned}
\end{equation}
with $\otimes$ denoting the Kronecker product. The equation in \eqref{eq:equation_of_interest} can be solved as a least-squares problem minimizing the Euclidean 2-norm $\| -\operatorname{vec}(u_{0: T-2}^{(1: M)}) - \mathscr{A}(z) \mathscr{D} \operatorname{vech}(Q) \|$ to obtain the value of the elements of the $Q$ matrix.

\begin{lemma} \label{lem:prop31}
If $M(T-2)m \geq N(N + 1)/2$ and $\mathscr{A}(z)\mathscr{D}$ has full column rank, then
the $Q \in S^n_+$ that corresponds to the given optimal trajectories $z_{0: T-1}^{(1: M)}$ is unique.
\end{lemma}
\begin{proof}
See related parts in \cite[Proposition 3.1]{discrete_ioc}.
\end{proof}

The first condition is related to having enough data so that \eqref{eq:equation_of_interest} is not under-determined and the second one to the controllability of $(A,B)$. Sometimes, $\mathscr{A}(z)\mathscr{D}$ does not have full column rank, but we can still get a unique $Q$ by the optimal trajectories $z_{0: T-1}^{(1: M)}$ available as shown in Lemma \ref{lem:thm31}.

\begin{lemma} \label{lem:thm31}
Suppose $T \geq N + 2$ and $M \geq N$. If there exists $N$ linearly independent $z_{T-1}^{(i)}$ among all $M$ sets of data, then there exists a unique $Q$ that corresponds to the given optimal trajectories $z_{0: T-1}^{(1: M)}$.
\end{lemma}
\begin{proof}
See related parts in \cite[Theorem 3.1]{discrete_ioc}.
\end{proof}

\section{Main results} \label{sec:main_results}
As shown in \cite{koopman_bil, strasser2025-overview-koopman}, the linear representation \eqref{eq:koopman_linear_disc} is less accurate for approximating a general nonlinear system than a bilinear one. Therefore, we focus on the following problem.

\begin{problem} \label{prob:main_prob}
Given optimal trajectory and control sequences $x^*_{0:T}, u^*_{0:T}$ for the unknown nonlinear system \eqref{eq:disc_dynamics} and objective function \eqref{eq:cost}, our goal is to estimate the corresponding bilinear representation of dynamics and LQR cost \eqref{eq:lqr_cost} state matrix estimate $\hat{Q}$ such that
\begin{equation} \label{eq:cost_equivalence}
    J_{\hat{Q}}(z^*_{0:T},u^*_{0:T}) \le J_{\hat{Q}}(z_{0:T},u_{0:T}),
\end{equation}
for all trajectories with a fixed starting point $z_0 = z^*_0$. Note that $J_{\hat{Q}} = J$ from \eqref{eq:lqr_cost} in the lifted dimensions $z$ with $R=I$.
\end{problem}


\subsection{Koopman-based separable bilinear system}
Based on the Koopman-based control approaches from \cite{Bevanda_2021}, we propose to use the optimal trajectories $(x^*_{0:T},u^*_{0:T})$ to obtain a bilinear control system in $z(x)$ and $u$ through a modified version of EDMDc. This allows us to linearize the dynamics with respect to the lifted state $z_k$ and capture nonlinear interactions through the bilinear terms. Since its control $u$ is the same as for the nonlinear case and the state does not depend on $u$, we can apply the PMP optimality conditions for this system to derive an inverse Bi-Linear Quadratic Regulator (Bi-LQR) formulation, which is inspired by the inverse LQR in \cite{discrete_ioc}, extended to the bilinear setting.

\subsubsection{Bilinear system representation}
We consider a bilinear representation in the lifted space as
\begin{equation}\label{eq:bilinear_dis} 
\begin{aligned} 
z_{k+1}=A z_k+B_1 z_k u_{1,k}+ \cdots +B_m z_k u_{m,k} = \\
A z_k+\left(\sum_{i=1}^m u_{i,k} B_i\right) z_k = 
A z_k+B\left(u_k \otimes z_k\right),
\end{aligned}
\end{equation}
\noindent where $B_i \in \mathbb{R}^{N \times N}$ for $i = 1, \dots, m$.

\begin{remark}\label{remark:unmodelled_dynamics}
If there exists a dictionary $\bar Z=\{z_i\}$ and a Koopman embedding $\theta:X \to \mathbb{R}^N$ with components in $\bar Z$ such that the span conditions of \cite[Thm.~II.1]{koopman_bil} hold, then the control–affine dynamics admit an exact bilinear Koopman realization \eqref{eq:bilinear_dis} and the approximation error $\epsilon_k:=\|\theta(x_{k+1})-(A\theta(x_k)+\sum_{i=1}^m u_{i,k}B_i\theta(x_k))\|$ is identically zero. Furthermore, the review paper \cite{strasser2025-overview-koopman} summarizes error bounds on the bilinear Koopman form, which can be made arbitrarily small based on the lifting dictionary, sample data size and distribution, and control input bounds.
\end{remark}


\subsubsection{Bilinear EDMDc} \label{sec:bi_edmdc}
We extend the results from Section \ref{sec:edmdc} to account for the bilinear dynamics \eqref{eq:bilinear_dis}. First, we stack all the data matrices using the $M$ optimal trajectories with $T$ data points each, i.e., $u_{0: T-1}^{(1: M)}$ and $x_{0: T-1}^{(1: M)}$:
\begin{equation}
    Z^+ = A Z + B \left[\begin{matrix}
        Z \odot \mathbbm{1}_N U_1 \
        \cdots \
        Z \odot \mathbbm{1}_N U_m
    \end{matrix}\right]^\top,
\end{equation}
where $Z = \theta(X) = [\theta(x_0^{(1: M)}), \theta(x_1^{(1: M)}), \dots, \theta(x_{T-1}^{(1: M)})]$, $U_i = [u_{i,0}^{(1: M)}, u_{i,1}^{(1: M)}, \dots, u_{i,T-1}^{(1: M)}]$ for $i = 1, \dots, m$, and $Z^+=\theta(Y) = [\theta(x_1^{(1: M)}), \theta(x_2^{(1: M)}), \dots, \theta(x_{T}^{(1: M)})]$. The superscript indicates stacked vectors of $M$ trajectory data, $\mathbbm{1}_N$ is a $N$-dimensional vector of ones and $\odot$ is the Hadamard product \cite{HornJohnson2012}. Matrices $A$ and $B$ can then be obtained by
\begin{equation}\label{eq:bi_koopman_approx}
    \min_{\left[\begin{array}{cc} A & B \end{array}\right]} 
    \left\| \:
    \theta(Y)-\left[
    \begin{array}{cc} A & B \end{array} 
    \right] \left[
    \begin{array}{c} \theta(X) \\ 
    \theta(X) \odot \mathbbm{1}_N U_1 \\
    \vdots \\
    \theta(X) \odot \mathbbm{1}_N U_m
    \end{array}
    \right] \right\|_F^2.
\end{equation}

As in Section \ref{sec:edmdc}, the closed form solution becomes
\begin{equation}\label{eq:bi_koopman_approx_cf}
    \left[\begin{array}{cc} A & B \end{array}\right] = \theta(Y) \left[ \begin{array}{c} \theta(X) \\ 
    \theta(X) \odot \mathbbm{1}_N U_1 \\
    \vdots \\
    \theta(X) \odot \mathbbm{1}_N U_m
    \end{array} \right]^\dag.
\end{equation}

\subsubsection{Bilinear inverse PMP}
We begin by considering the dynamics \eqref{eq:bilinear_dis} and apply PMP to derive necessary conditions for optimality. The Hamiltonian function is then defined as
\begin{equation} \label{eq:hamiltonian_bipmp} 
\begin{aligned}
H(z_k, u_k, \lambda_{k+1}) = & \lambda_{k+1}^\top \left( A z_k + \sum_{i=1}^m B_i z_k u_{i,k} \right) + \\ 
& \frac{1}{2} z_k^\top Q z_k + \frac{1}{2} u_k^\top R u_k, 
\end{aligned}
\end{equation} 
where $R = I$ for simplicity, $u_k = [u_{1,k}, u_{2,k}, \dots, u_{m,k}]^\top$, and $\lambda_{k+1}$ is the co-state at time $k+1$. We also define the following matrices, for each time step $k$, to simplify our notation:
\begin{equation} \label{eq:O_AB} O_{AB_k} = A + \sum_{i=1}^m B_i \otimes u_{i,k} \in \mathbb{R}^{N \times N}, 
\end{equation} 
\begin{equation} \label{eq:O_B} O_{B_k} = \left[ \begin{array}{cccc} B_1 z_k & B_2 z_k & \cdots &  B_m z_k\end{array} \right] \in \mathbb{R}^{N \times m}. 
\end{equation}

With these, the PMP conditions become more tractable. The derivative of the Hamiltonian with respect to state $z_k$ yields
\begin{equation} \label{eq:bi_PMP_condition_lambda} 
\lambda_{k} = \frac{\partial H}{\partial z_k} = Q z_k + O_{AB_k}^\top \lambda_{k+1}, \quad \text{for } k = 1, \dots, T-1,
\end{equation} 
where the terminal condition is $\lambda_{T} = 0$. The condition on the derivative of the Hamiltonian with the control input $u_k$ gives
\begin{equation} \label{eq:bi_pmp_optimal_u} 
u_k = - O_{B_k}^\top \lambda_{k+1}, \quad \text{for } k = 0, \dots, T-1. 
\end{equation}

We now introduce the main theorem that relates the control inputs, the co-states, and the unknown cost matrix $Q$.

\begin{theorem} \label{thm:bilinear_inverse_pmp} 
Given the bilinear system dynamics \eqref{eq:bilinear_dis} and the Hamiltonian function \eqref{eq:hamiltonian_bipmp}, the necessary conditions of optimality lead to the following relationship between the vectorized control inputs and the unknown cost matrix $Q$
\begin{equation} \label{eq:bilinear_pmp_equation}
-\operatorname{vec}\left(u_{0: T-2}^{(1:M)}\right) = \mathcal{A}(z, u) \operatorname{vec}(Q) = \mathcal{A}(z, u) \mathscr{D} \operatorname{vech}(Q), 
\end{equation}
where $\mathcal{A}(z, u)$ is defined as
\begin{equation} \label{eq:big_A_stacked}
\mathcal{A}\left(z,u\right) = \left[ 
\begin{array}{cccc} \mathcal{A}_1\left(z,u\right) & \mathcal{A}_2\left(z,u\right) & \cdots & \mathcal{A}_M\left(z,u\right)
\end{array} \right]^\top,
\end{equation}
and $\mathcal{A}_i\left(z,u\right)$ being
\begin{equation} \label{eq:big_A}
\begin{aligned}
\mathcal{A}_i\left(z,u\right) = & \sum_{k=1}^{T-1} z_k^{(i)^\top} \otimes 
\left[\begin{array}{c} 
\Bigg( \prod_{l=1}^{k-1} O_{AB_l}^{(i)^\top} \Bigg) O_{B_0}^{(i)^\top} \\
\Bigg( \prod_{l=2}^{k-1} O_{AB_l}^{(i)^\top} \Bigg) O_{B_1}^{(i)^\top} \\
\vdots \\
\Bigg( \prod_{l=T-1}^{k-1} O_{AB_l}^{(i)^\top} \Bigg) {O_B}_{T-2}^{(i)^\top}
\end{array} \right]
\end{aligned}
\end{equation}
with the convention that the product over an empty index set is the identity matrix.
\end{theorem}

\begin{proof} 
We follow the same process as in Section \ref{sec:ilqr}. We start by stacking the PMP condition \eqref{eq:bi_PMP_condition_lambda} across all $M$ trajectories with $T$ points each. Hence, for trajectory $i$, we obtain $\mathcal{O}_{AB}[\begin{array}{ccc}
\lambda_1^{(i)} &
\cdots &
\lambda_T^{(i)}
\end{array}]^\top =
[\begin{array}{cccc}
Q z_1^{(i)} &
\cdots &
Q z_{T-1}^{(i)} &
0
\end{array}]^\top$ 
\begin{equation} \label{eq:lambda_stacked}
\mathcal{O}_{AB} = \left[\begin{array}{cccc}
I & -{O_{AB}}_1^{(i)^\top} &  & \\
& I & \ddots & \\
& & \ddots & -{O_{AB}}_{T-1}^{(i)^\top} \\
& & & I
\end{array}\right],
\end{equation}
where the blank entries in the matrix are zeros. By recursively solving the above system, we express the co-states $\lambda_k^{(i)}$ in terms of the state variables $z_k^{(i)}$
\begin{equation} \label{eq:lambda_recursive} 
\lambda_k^{(i)} = Q z_k^{(i)} + \sum_{j=k}^{T-2} \left( \prod_{l=k}^{j} O_{AB_l}^{(i)^\top} \right) Q z_{j+1}^{(i)}, \: k = 1, \dots, T-1.
\end{equation} 

Next, we stack the optimal control condition \eqref{eq:bi_pmp_optimal_u} to obtain 
\begin{equation*}
\begin{aligned} 
-\operatorname{vec}&\left(u_{0: T-2}^{(i)}\right) = \\ & diag([{O_B}_0^{(i)^{\top}}, {O_B}_1^{(i)^{\top}},\dots,\ {O_B}_{T-1}^{(i)^{\top}}] \operatorname{vec}\left(\lambda_{1: T-1}^{(i)}\right).
\end{aligned}
\end{equation*}

Substituting \eqref{eq:lambda_recursive} in $\operatorname{vec}\left(\lambda_{1: T-1}^{(i)}\right)$, we obtain
\begin{equation} \label{eq:u_vs_Q}
-\operatorname{vec}\left(u_{0: T-2}^{(i)}\right) = \mathcal{A}_i(z, u) \operatorname{vec}(Q), 
\end{equation} 
where $\mathcal{A}_i(z, u)$ is computed using \eqref{eq:big_A}. Finally, stacking the equations for all $M$ trajectories, we get \eqref{eq:big_A_stacked} and therefore the condition in $Q$ shown in \eqref{eq:bilinear_pmp_equation}.
\end{proof}

As in Section \ref{sec:ilqr}, the well-, or over-determined equations in \eqref{eq:bilinear_pmp_equation} can be solved as a least-squares problem minimizing the Euclidean 2-norm $\| -\operatorname{vec}(u_{0: T-2}^{(1: M)}) - \mathcal{A}(z,u) \mathscr{D} \operatorname{vech}(Q) \|$ to obtain the value of the elements of the $Q$ matrix.

\begin{lemma} \label{lem:uniqueness_condition} 
If $M(T-2)m \geq N(N + 1)/2$ and $\mathcal{A}(z, u)$ has full column rank, then the unknown matrix $Q$ corresponding to the given optimal trajectories $z_{0: T-2}^{(1: M)}$ is unique. 
\end{lemma}

\begin{proof} 
The number of equations in \eqref{eq:bilinear_pmp_equation} is $M(T-2)m$, and the number of unknowns in $\operatorname{vech}(Q)$ is $N(N + 1)/2$. If $M(T-2)m \geq N(N + 1)/2$ and $\mathcal{A}(z, u)$ has full column rank, \eqref{eq:u_vs_Q} has a unique solution for $\operatorname{vech}(Q)$. 
\end{proof}

As in Section \ref{sec:ilqr}, $\mathcal{A}\left(z,u\right)\mathscr{D}$ does not necessarily need to be full column rank, but we can still get a unique $Q$ by the optimal trajectories $z_{0: T-1}^{(1: M)}$ available as shown in Lemma \ref{lem:linearly_independent_states}.

\begin{lemma} \label{lem:linearly_independent_states} 
Suppose $T \geq N + 2$ and $M \geq N$. If there exist $N$ linearly independent $z_{T-1}^{(i)}$ among the $M$ trajectories, then the $Q$ corresponding to the given optimal trajectories is unique. 
\end{lemma}

\begin{proof} 
The existence of $N$ linearly independent terminal states in $z_{T-1}^{(1:M)}$ ensures that the matrix $\mathcal{A}(z, u)$ spans the necessary space to uniquely determine $Q$. This condition guarantees that the data is sufficiently rich and that the controllability of the system contributes to the full rank of $\mathcal{A}(z, u)$. The proof is trivial since it is equivalent to that of Lemma \ref{lem:thm31}, adapted to the bilinear dynamics \eqref{eq:bilinear_dis}. 
\end{proof}

These lemmas formalize the conditions under which the cost matrix $Q$ can be uniquely identified from the observed optimal trajectories. The first lemma emphasizes the need for enough data to avoid an underdetermined system, while the second highlights the importance of state diversity among trajectories.

\subsection{Proposed solution}
Algorithm~\ref{alg:IOC_koopman} solves Problem~\ref{prob:main_prob} through: (i) data collection (step~$1$) of optimal state and control trajectories $x_{0:T-1}^{(1:M)},u_{0:T-1}^{(1:M)}$. (ii) Koopman lifting (steps~$2-3$) via bilinear EDMDc \eqref{eq:bi_koopman_approx_cf} to approximate~\eqref{eq:bilinear_dis}. (iii) Inverse Bi-LQR (steps~$4-6$) to estimate cost matrices with bilinear PMP \eqref{eq:bilinear_pmp_equation}.

\begin{algorithm}[ht]
\caption{Koopman-based Inverse Bi-LQR framework}
\label{alg:IOC_koopman}
\begin{algorithmic}[1]
\STATE \textbf{Initialization} Obtain optimal trajectory tuples $(x_k, u_k, x_{k+1})$ for $M$ trajectories, where $x_{k+1}$ follows \eqref{eq:disc_dynamics} for $k = 0, 1, ..,T-1$ and stack them accordingly as $u_{0: T-1}^{(1: M)}$, $x_{0: T-1}^{(1: M)}$ and $x_{1: T}^{(1: M)}$.
\STATE \textbf{EDMDc} Create vector of basis functions $\theta(x) = [\theta_1(x), \theta_2(x), ..., \theta_N(x)]^\top$ and compute matrices $\theta(X) = [\theta(x_0^{(1: M)}), \theta(x_1^{(1: M)}), \dots, \theta(x_{T-1}^{(1: M)})]$, $U_i = [u_{i,0}^{(1: M)}, u_{i,1}^{(1: M)}, \dots, u_{i,T-1}^{(1: M)}]$ for $i = 1, \dots, m$, and $\theta(Y) = [\theta(x_1^{(1: M)}), \theta(x_2^{(1: M)}), \dots, \theta(x_{T}^{(1: M)})]$.
\STATE \textbf{EDMDc} Compute $A$ and $B$ using \eqref{eq:bi_koopman_approx_cf}.
\STATE \textbf{Bi-LQR} Calculate ${O_{AB}}_k^{(i)^{\top}}$ and ${O_B}_k^{(i)^{\top}}$ for $k=0,\dots,T-1$ and $i=1,\dots,M$ using \eqref{eq:O_AB} and \eqref{eq:O_B}.
\STATE \textbf{Bi-LQR} Obtain the elements $\mathcal{A}_i\left(z,u\right)$ for $i=1,\dots,M$ using \eqref{eq:big_A} and compute the final stacked data matrix $\mathcal{A}\left(z,u\right)$ using \eqref{eq:big_A_stacked}.
\STATE \textbf{Bi-LQR} Lastly, solve for $Q$ by minimizing 
\[
\| -\operatorname{vec}(u_{0: T-2}^{(1: M)}) - \mathcal{A}(z,u) \mathscr{D} \operatorname{vech}(Q) \|,
\]
\Return{$A$, $B$, $Q$.}
\end{algorithmic}
\end{algorithm}

With $\Delta t$ chosen so that $\Delta t^2 \ll \Delta t<1$ and using enough data, the bilinear EDMDc model $(A,B)$ approximates \eqref{eq:disc_dynamics} with minimal error $\epsilon_k$. Additionally, by Theorem \ref{thm:bilinear_inverse_pmp} and Lemma \ref{lem:uniqueness_condition} or Lemma \ref{lem:linearly_independent_states}, the recovered cost $Q$ will be unique and optimal for the original trajectories, $(x^*_{0:T},u^*_{0:T})$, thanks to the application of \eqref{eq:equation_of_interest}. Therefore, \eqref{eq:cost_equivalence} will hold.

\begin{remark}
Since LQR solvers cannot be used with the bilinear dynamics in \eqref{eq:bilinear_dis}, the solution employed for this work is to transform the bilinear dynamics back to nonlinear \eqref{eq:disc_dynamics} as
$
x_{k+1}=C A \theta\left(x_k\right) + \sum_{i=1}^m C B_i \theta\left(x_k\right) u_i,
$
where matrix $C$ is computed using the EDMDc algorithm from the bilinear dynamics to the nonlinear ones with 
$
\min_{C} \left\| \: X - C \theta(X) \right\|_F^2,
$
which can be readily solved by $C = X \theta(X)^\dag$. The cost can then be formulated as 
$
J_{B i}=\theta(x)^{\top} Q \theta(x)+u^{\top} R u.
$
\end{remark}

\section{Simulations}

\begin{figure*}[t]
    \centering
    \begin{subfigure}[t]{0.39\textwidth}
        \centering
        \includegraphics[width=\linewidth]{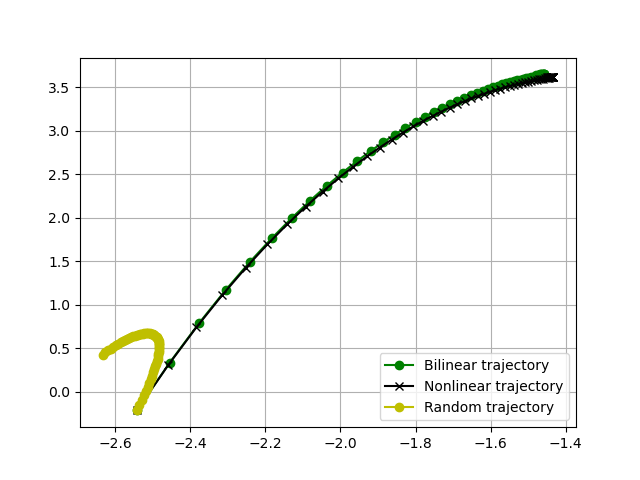}
        \subcaption{\centering Evolution of the system for Example~\ref{ex:controllable}.}
        \label{fig:example_2}
    \end{subfigure}\hfill
    \begin{subfigure}[t]{0.39\textwidth}
        \centering
        \includegraphics[width=\linewidth]{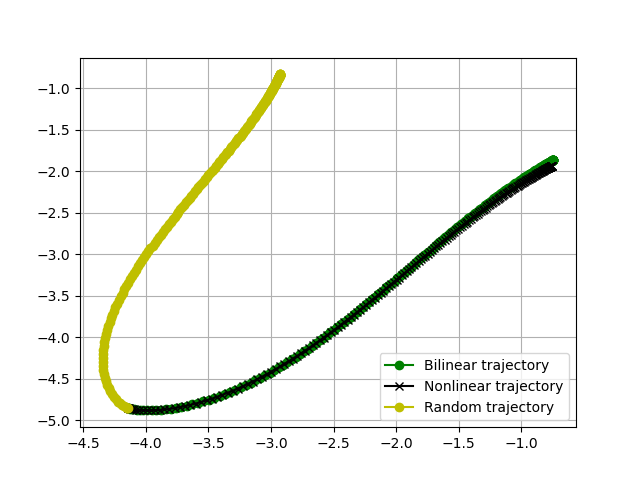}
        \subcaption{\centering Evolution of the simulated robot teleoperated \\ by a human.}
        \label{fig:example_human}
    \end{subfigure}\hfill
    \begin{subfigure}[t]{0.2\textwidth}
        \centering
        \includegraphics[width=\linewidth]{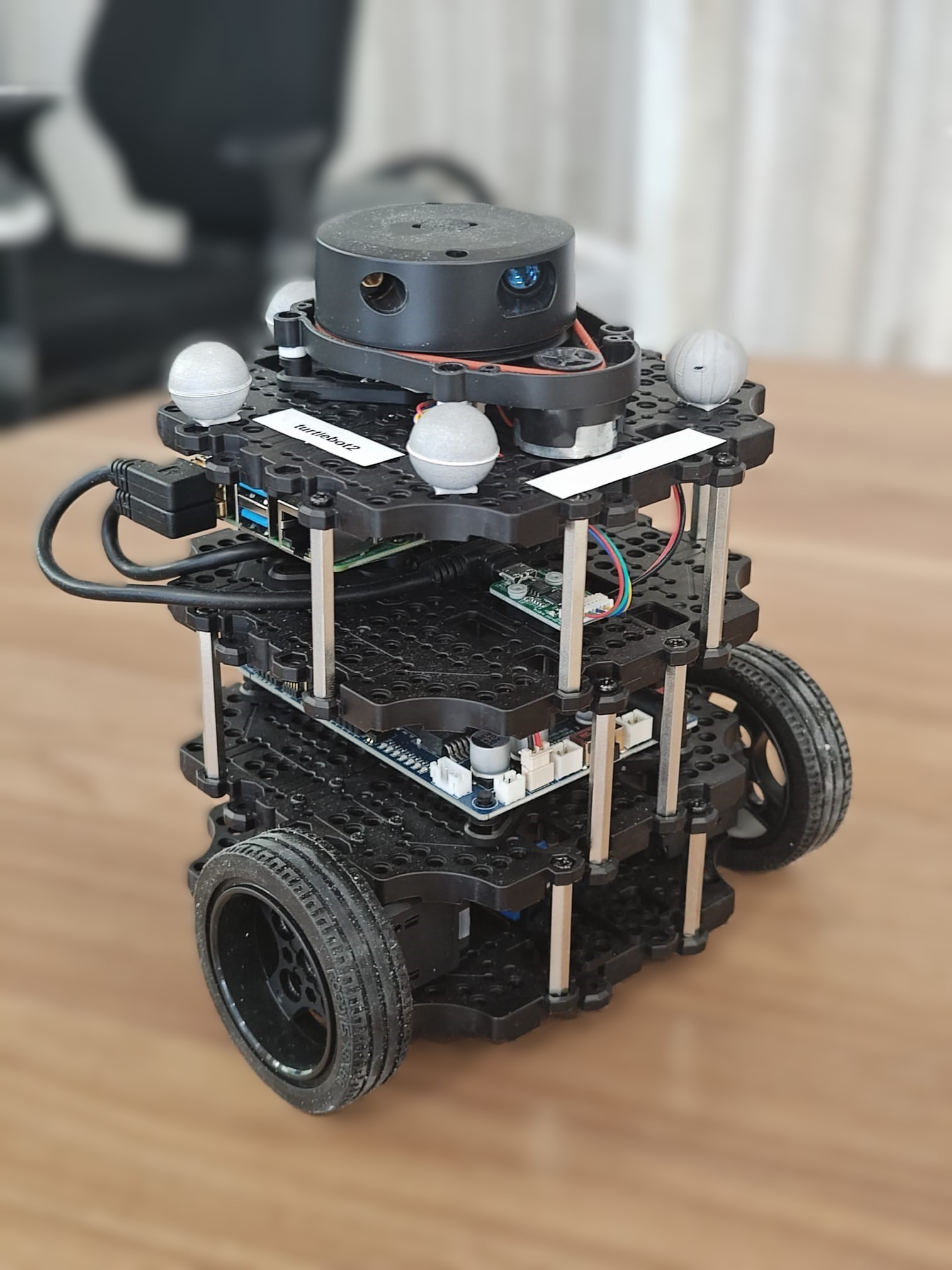}
        \subcaption{\centering The TurtleBot3 used in the experiments.}
        \label{fig:turtlebot}
    \end{subfigure}
    \caption{Side-by-side comparison of simulation and hardware setup.}
    \label{fig:three_side_by_side}
\end{figure*}

We now present simulations that illustrate the method on a simple example and on a teleoperated robot. To solve the optimization problem, we use off-the-shelf nonlinear solvers, e.g., CasADi \cite{Andersson2019}, but a more efficient approach would be to use bilinear LQR solvers \cite{ensemble_bilqr}.

\subsection{Controllable example}
\begin{example} \label{ex:controllable}
Consider the 2-d discrete time control system
\begin{equation} \label{eq:ex_2_dis}
\left[ \begin{array}{c}
     x_{1,k+1}  \\
     x_{2,k+1}  
 \end{array}  \right]  = \left[ \begin{array}{c}
     (1+c \Delta t) x_{1,k}+\Delta t u_{1,k}  \\
     (1+d \Delta t) x_{2,k} + (d-2c)\Delta t x_{1,k}^2 + \\ \Delta t x_{1,k}^2 u_{1,k} + \Delta t u_{2,k}
 \end{array}  \right],
\end{equation}
where $c,d \in[0,1]$. The system is stable at the origin, and has invariant manifolds $x_1 = 0$ and $x_2 = -x_1^2$. In the lifted state space $z = \theta(x) = [x_1, x_2 + x_1^2, x_1^2, 1]^\top$, the dynamics become bilinear as follows
\begin{align} \label{eq:ex_2_bidis}
z_{k+1} & = \left[\begin{array}{cccc}
1+c \Delta t & 0 & 0 & 0 \\
0 & 1+d \Delta t & c^2 \Delta t^2 & 0 \\
0 & 0 & 1+2 c \Delta t+c^2 \Delta t^2 & 0 \\
0 & 0 & 0 & 1
\end{array}\right] z_k \notag \\
& + \left[\begin{array}{cccc}
0 & 0 & 0 & \Delta t \\
2 \Delta t+c \Delta t^2 & 0 & \Delta t & 0 \\
2 \Delta t+c \Delta t^2 & 0 & 0 & 0 \\
0 & 0 & 0 & 0
\end{array}\right] z_k u_{1,k} 
\nonumber 
\displaybreak[2]\\
& + \left[\begin{array}{cccc}
0 & 0 & 0 & 0 \\
0 & 0 & 0 & \Delta t \\
0 & 0 & 0 & 0 \\
0 & 0 & 0 & 0
\end{array}\right] z_k u_{2,k}+\left[\begin{array}{c}
0 \\
\Delta t^2 u_{1,k}^2 \\
\Delta t^2 u_{1,k}^2 \\
0
\end{array}\right].
\end{align}
\end{example}

Using the basis $\phi = [\phi_1,\dots,\phi_N] = [x_1^2, x_2^2, x_1^4, 1, u_1^2, u_2^2]$ with weights $w = [w_1,\dots,w_N] = [1, 2, 3, 1, 1, 1]$, we generate optimal trajectories which are then inputted to Algorithm $\ref{alg:IOC_koopman}$, resulting in the following
\begin{equation*}
\begin{aligned}
diag(A) = [1.0030, 1.0020, 1.0060, 1],
\end{aligned}
\end{equation*}
\begin{equation*}
\begin{aligned}
diag(Q) = [0.6870, 1.8783, 2.8183, 0],
\end{aligned}
\end{equation*}
with the off-diagonal elements being negligible or zero, and
\begin{equation*}
B_1 = \left[ \begin{matrix}
    0 & 0 & 0 & 0.01 \\
    0 & 0 & 0.01 & 0 \\
    0.02 & 0 & 0 & 0 \\
    0 & 0 & 0 & 0 
    \end{matrix}\right], \
B_2 = \left[ \begin{array}{c|c}
\multicolumn{1}{c}{0_{4\times 3}} & \begin{matrix}
0 \\
0.01 \\
0 \\
0
\end{matrix} \end{array} \right],
\end{equation*}
where $0_{n\times m}$ is a $m\times n$ matrix of all zeros.

Because $\Delta t=0.01$ satisfies $\Delta t^2\!\ll\!\Delta t<1$, the squared terms in~\eqref{eq:ex_2_bidis} can be neglected, yielding an accurate bilinear approximation of~\eqref{eq:ex_2_dis}. Smaller $\Delta t$ increases accuracy, but also the required data and computation time. Fig.~\ref{fig:example_2} shows the system evolution and an optimal trajectory using randomly selected weights $w$ to show the problem is not trivial.

\begin{figure*}[t]
    \centering
    \begin{minipage}[b]{0.49\textwidth}
        \centering
        \includegraphics[width=\linewidth]{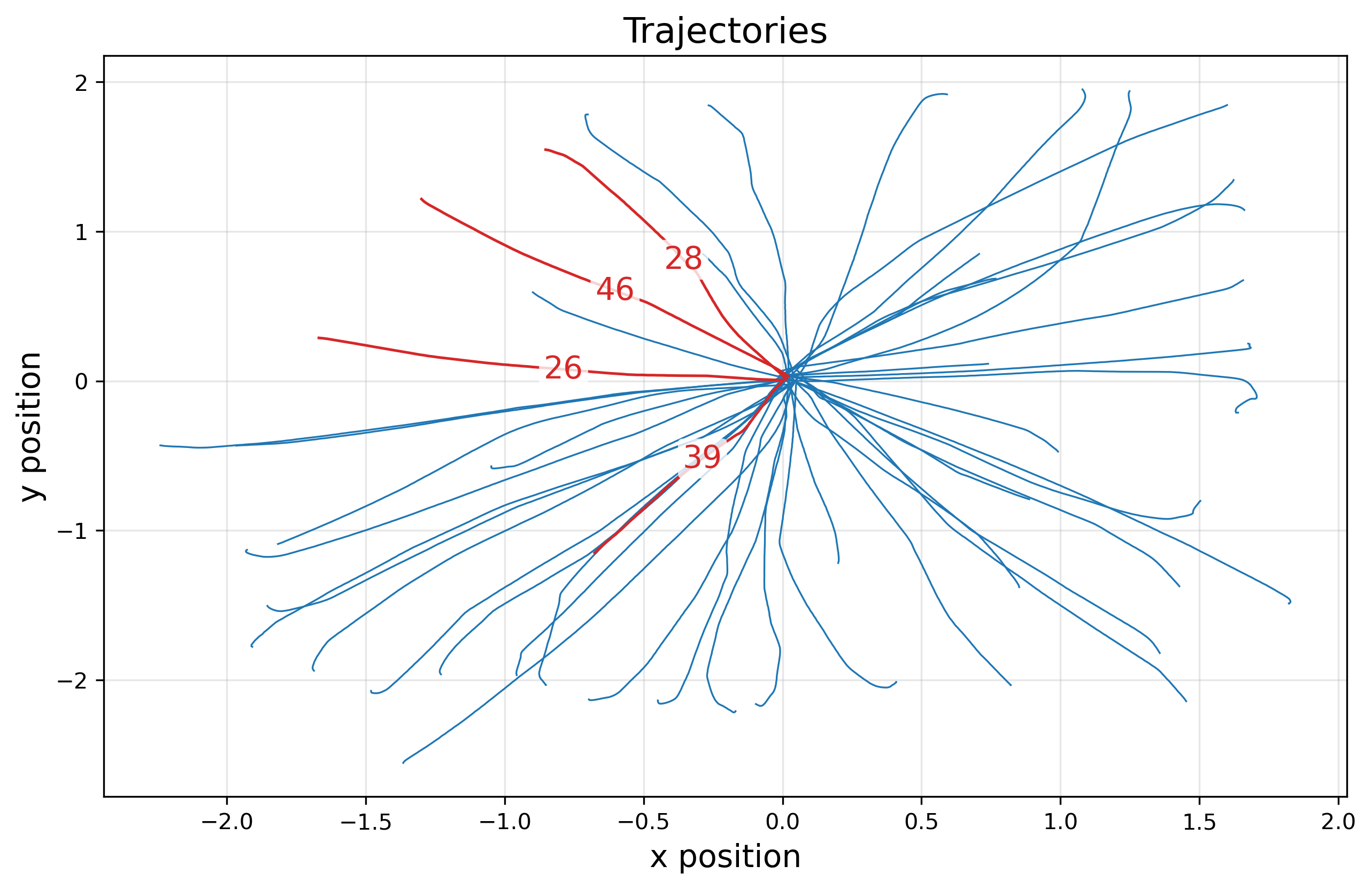}
        \subcaption{Recorded trajectories with training (blue) and test (red) runs.}
        \label{fig:trajs_all}
    \end{minipage}
    \hfill
    \begin{minipage}[b]{0.49\textwidth}
        \centering
        \subcaptionbox{Trajectory 26\label{fig:traj26}}%
            {\includegraphics[width=0.49\linewidth]{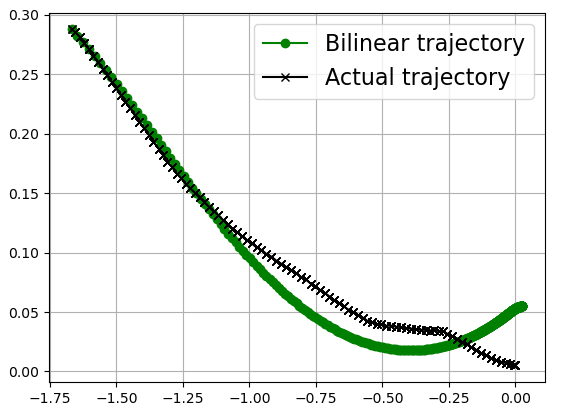}}
        \subcaptionbox{Trajectory 28\label{fig:traj28}}%
            {\includegraphics[width=0.49\linewidth]{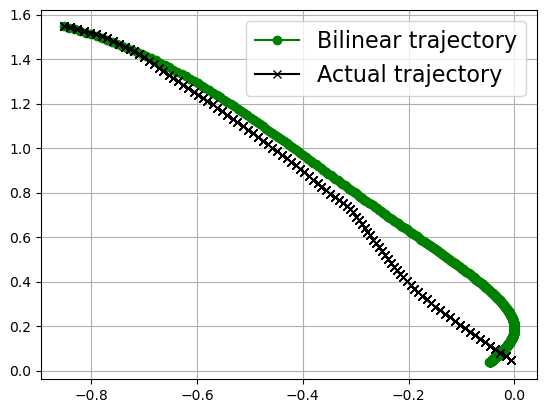}}\\[0.5ex]
        \subcaptionbox{Trajectory 39\label{fig:traj39}}%
            {\includegraphics[width=0.49\linewidth]{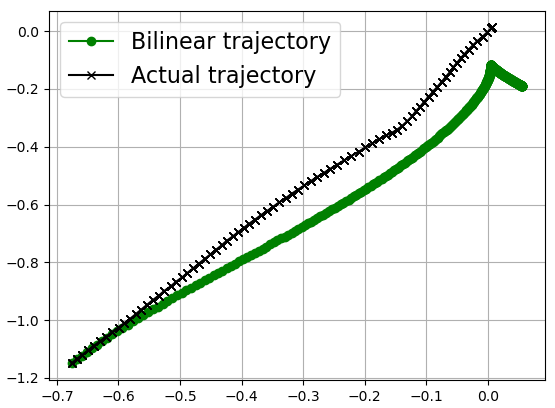}}
        \subcaptionbox{Trajectory 46\label{fig:traj46}}%
            {\includegraphics[width=0.49\linewidth]{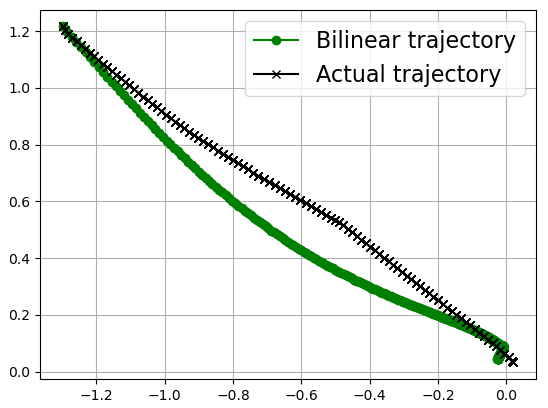}}
    \end{minipage}

    \caption{(a) All trajectories. (b, c, d, e) Actual (black) vs.\ predicted trajectory (green) comparisons for the four test cases.}
    \label{fig:all_experiments}
\end{figure*}

\subsection{Estimating human cost and robot dynamics} \label{sec:sim_human}
To illustrate our original point of human prediction, we created a 2D simulation where a ``fictitious" human remotely teleoperates a robot modeled by unknown unicycle dynamics
\begin{equation*}
x_{k+1}=\left[\begin{array}{l}
x_{1,k+1} \\
x_{2,k+1} \\
x_{3,k+1}
\end{array}\right]=\left[\begin{array}{l}
x_{1,k}+u_{1,k} \cos \left(x_{3,k}\right) \Delta t \\
x_{2,k}+u_{1,k} \sin \left(x_{3,k}\right) \Delta t \\
x_{3,k}+u_{2,k} \Delta t
\end{array}\right].
\end{equation*}

In the lifted state space 
\[
z=\theta(x)=\left[\begin{array}{cccccc}
x_1 & x_2 & x_3 & \cos \left(x_3\right) & \sin \left(x_3\right) & 1
\end{array}\right],
\]
\begin{equation*}
\begin{aligned}
z_{k+1} = I_{6\times 6} z_k & +
\left[\begin{array}{cccccc}
0 & 0 & 0 & \Delta t & 0 & 0 \\
0 & 0 & 0 & 0 & \Delta t & 0 \\
\multicolumn{6}{c}{0_{4\times 6}}
\end{array}\right] z_k u_{1,k} \\
& +
\left[\begin{array}{cccccc}
\multicolumn{6}{c}{0_{2\times 6}} \\
0 & 0 & 0 & 0 & 0 & \Delta t \\
0 & 0 & 0 & 0 & -\Delta t & 0 \\
0 & 0 & 0 & \Delta t & 0 & 0 \\
0 & 0 & 0 & 0 & 0 & 0
\end{array}\right] z_k u_{2,k},
\end{aligned}
\end{equation*}
where $I_{6\times6}$ is the 6-dimensional identity matrix. The objective of the ``fictitious" human is modeled using basis $\phi = [\phi_1,\dots,\phi_N] = [x_1^2, x_2^2, x_3^2, \cos(x_3)^2, \sin(x_3)^2, 1, u_1^2, u_2^2]$ with weights $w = [w_1,\dots,w_N] = [1, 1, 1, 0, 0, 0, 1, 1]$. The results of Algorithm $\ref{alg:IOC_koopman}$ were the following
\begin{equation*}
\begin{aligned}
diag(A) = [1, 1, 1, 0.9997, 1.0007, 1],
\end{aligned}
\end{equation*}
\begin{equation*}
\begin{aligned}
diag(Q) = [1.0305, 1.0306, 1.0485, 4.4317, 4.2081, 0],
\end{aligned}
\end{equation*}
with the off-diagonal elements being negligible or zero, and
\begin{equation*}
B_1 =
\begin{bmatrix}
0 & 0 & 0 & 0.01 & 0 & 0 \\
0 & 0 & 0 & 0    & 0.01 & 0 \\
\multicolumn{6}{c}{0_{4\times 6}}
\end{bmatrix},
\end{equation*}
\begin{equation*}
B_2 =
\begin{bmatrix}
\multicolumn{6}{c}{0_{2\times 6}} \\
0 & 0 & 0 & 0 & 0 & 0.01 \\
0 & 0 & 0 & 0 & -0.01 & 0 \\
0 & 0 & 0 & 0.01 & 0 & 0 \\
0 & 0 & 0 & 0 & 0 & 0
\end{bmatrix}.
\end{equation*}

As before, since the sampling time used here was $\Delta t = 0.01 \implies \Delta t^2 \ll \Delta t<1$, we conclude that the discrete-time form can be used to accurately model the original nonlinear dynamics. The evolution of the system is shown in Fig. \ref{fig:example_human}.

\section{Experiment}
To validate our framework in a real-world setting, we conducted experiments using a TurtleBot3 (see Fig.~\ref{fig:turtlebot}). The setup was the same as in Section~\ref{sec:sim_human}, with the robot's position and velocity recorded using the Qualisys Motion Capture System. The arena was a flat surface free of obstacles. 48 trajectories were collected (see Fig.~\ref{fig:trajs_all}), varying in length and complexity. Of these, 44 were used for training and 4 (numbers 26, 28, 39, and 46) were randomly selected for testing.

We then applied Alg.~\ref{alg:IOC_koopman} with the same basis functions as in Sec.~\ref{sec:sim_human}. The results (Figs.~\ref{fig:traj26} to \ref{fig:traj46}) show the predicted trajectories closely follow the actual ones, capturing the dynamics and control employed by the human operator. While minor discrepancies arise due to unmodeled dynamics and noise, the accuracy is sufficient for practical applications. In particular, the model enables short-horizon predictions, valuable for downstream tasks, e.g., collision avoidance.

\section{Conclusion}

We proposed a Koopman-based Inverse Optimal Control framework that learns a bilinear model from demonstrations via a modified EDMDc and recovers an LQR-type cost in the lifted space using PMP conditions akin to inverse LQR. This avoids nonlinear IOC complexity while handling unknown nonlinear dynamics. Simulations and a TurtleBot experiment demonstrate accurate recovery of dynamics and cost, useful for short-horizon prediction. Future work will target higher-dimensional tasks, richer dictionaries (e.g., RBF/Fourier), robustness to noise and dedicated faster bilinear LQR solvers.

\section*{References}
\bibliographystyle{ieeetr}
\bibliography{Reference}

\end{document}